# Mid infrared imaging of mass transport in polymer electrolyte membranes of an operating microfluidic water electrolyzer

Stéphane CHEVALIER[1,2,*], Meguya RYU[3], Jean-Christophe BATSALE[2], Junko MORIKAWA[4,5]

E-mail: schevali@iis.u-tokyo.ac.jp, morikawa.j.4f50@m.isct.ac.jp

[1] LIMMS/CNRS-IIS(IRL2820), Institute of Industrial Science, The University of Tokyo, Japan

[2] Arts et Metiers Institute of Technology, CNRS, Bordeaux INP, I2M, UMR 5295, F-33400 Talence, France

[3] National Metrology Institute of Japan (NMIJ), National Institute of Advanced Industrial Science and Technology (AIST), Tsukuba 305-8563, Japan

[4] School of Materials and Chemical Technology, Institute of Science Tokyo (Science Tokyo), Meguro-ku, Tokyo 152-8550, Japan

[5] Research Center for Autonomous Systems Materialogy (ASMat), Institute of Integrated Research (IIR), Institute of Science Tokyo (Science Tokyo), Yokohama 226-8501, Japan

This study investigates water transport in a polymer electrolyte membrane (PEM) electrolyzer using operando infrared spectroscopic imaging. By testing different $H_2SO_4$ anolyte concentrations, it examines electrochemical performance, water diffusion, and membrane hydration. Higher anolyte concentrations increased standard deviations in current densities and led to water diffusion gradients revealed by infrared imaging and confirming localized water transport variations. The study highlights the need for improved water management and optimized electrolyzer design for stable and efficient PEM electrolysis in industrial applications.

……………………………………………………………………

1. Introduction

Over the past couple decades, poorly regulated anthropogenic activities have led to rapidly increasing global temperatures and have triggered the sixth and ongoing extinction event – The Holocene Extinction [1]. This looming threat of climate change has motivated the shift in energy usage from fossil fuel consumption to renewable energy. While the shift to renewable energy is promising and is scaling up globally, it must also accommodate the increasing global energy consumption which was forecasted by the EIA to increase by 56 % between 2010 and 2040 [2]. In addition to the development of renewable energy technologies such as wind and solar, energy storage infrastructure must be developed in tandem to compensate for the intermittent nature of these renewable technologies [3,4].

A hydrogen economy has been proposed as a potential solution for energy storage when coupled with intermittent renewable energy technologies [5,6]. Electrical energy produced through wind and solar can be converted and stored to chemical energy in the form of hydrogen as an energy carrier. Polymer electrolyte membrane (PEM) water electrolyzers are one of the most promising technologies for fulfilling this necessity, through the electrochemical splitting of water in the absence of carbon emissions. Moreover, PEM electrolysis benefits from: high purity hydrogen production (>99.95 %), fast system responses (on the millisecond scale), high operating current densities, and flexible operating conditions [7,8]. However, the high capital costs and low efficiencies of these devices have slowed their commercialization, and are consequently key research targets for spearheading this technology [8].

Thus, optimizing the efficiency of PEM water electrolysis is imperative for achieving the future hydrogen economy. Understanding and improving mass and ionic transport mechanisms within the membrane used in polymer electrolyte membrane (PEM) water splitting electrolyzers is vital for achieving improved efficiencies that would enable the use of water electrolysis in sustainable energy infrastructures. In this work, a novel microfluidic water electrolysis chip with an integrated PEM was developed for characterization via operando infrared (IR) spectroscopy [9,10].

This study explores the transport mechanisms and loss sources in a polymer electrolyte membrane (PEM) electrolyzer using a combination of electrochemical and imaging techniques. The electrolyzer was tested under different $H_2SO_4$ anolyte concentrations to examine their impact on electrochemical performance, water diffusion, and membrane hydration. Infrared (IR) spectroscopic imaging was used to visualize water transport within the PEM, revealing that higher anolyte concentrations led to increased standard deviations in current densities and the development of water diffusion gradients from anode to cathode. These gradients impact the overall efficiency and stability of the electrolyzer, necessitating improved water management strategies.

2. Methods

2.1. Experimental setup

An in-house IR spectroscopy setup (Fig 1) was developed for operando imaging of the PEM electrolyzer. Light was produced from a black body source with a temperature of 1400 K. The emitted light was collimated (Thorlabs ID75Z/M) and then filtered to a wavelength of 4 µm (wavenumber of 2500 $cm^{-1}$) using a band pass filter. The filtered beam was collimated again (Thorlabs ID25/M) before being attenuated by the sample, and then captured by an IR camera (FLIR SC7000 with InSb sensor). Images were captured with a spatial resolution of 21.6 µm $px^{-1}$.

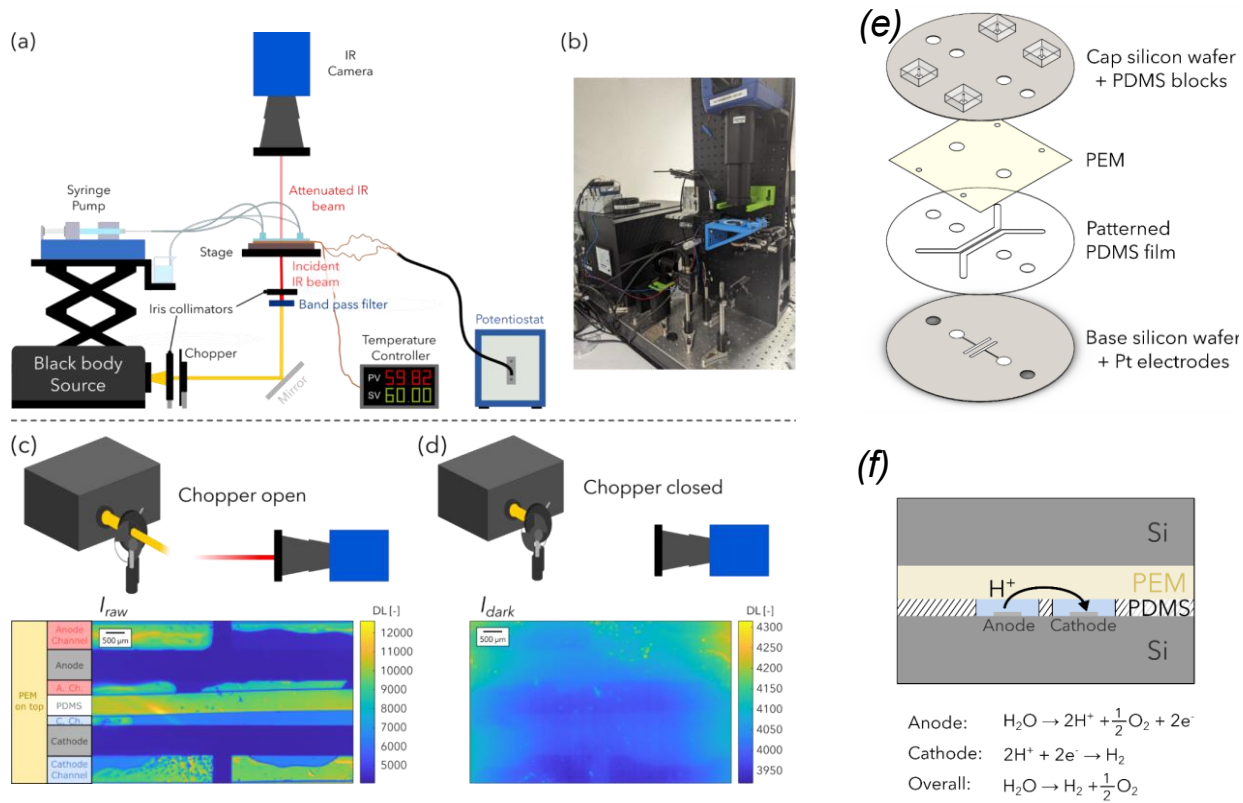

**Figure 1. (a)** *(a) Schematic of the experimental setup used. (b) Photo of the experimental setup without the microfluidic electrolyzer, potentiostat, temperature controller, and syringe pump. When the chopper is (c) open, raw images $I_{raw}$ are captured, and when the chopper is (d) closed, dark images $I_{dark}$ are captured. (e) schematic of the PEM water electrolyzer used in this study. (f) Cut view of the PEM water electrolyzer with the proton path described.*

A computer was connected to the IR camera and a chopper positioned between the first collimator and the band pass filter to align the frequency of the chopper to the camera's frame rate. The synchronization between the chopper and the IR camera was configured such that captured images would alternate between raw images ($I_{raw}$) and dark-field images ($I_{dark}$), as shown in Fig 1c and **Erreur ! Source du renvoi introuvable.**d. Raw images Fig 1c show the features of the microfluidic electrolyzer, where the PEM covers the entire image, each channel can be seen, as well as the respective electrodes that block the beam. Additionally, liquid reactants and gaseous products can be visually differentiated through differences in the image's digital level (DL). A higher concentration electrolyte flowed through the anode channel for the sample image shown Fig 1c and **Erreur ! Source du renvoi introuvable.**d, which attenuated the beam more than the electrolyte that flowed through the cathode channel. A schematic of the microfluidic electrolyzer is also presented in Fig. 1(e) and (f). More detailed information about it can be found in [9].

2.2. Experimental protocol

The water content in the PEM of the microfluidic electrolyzer was determined for acidic water splitting using diluted $H_2SO_4$ electrolyte while IR images were concurrently acquired. For acidic water splitting, protons (H+ ions) are consumed at the cathode (Eq. 1). Supplying higher anolyte $H_2SO_4$ concentrations was hypothesized to saturate the H+ ion availability in the anolyte channel. Electro-osmotic drag is then assumed to occur at an accelerated rate, which would ideally manifest as water gradients from the anode to the cathode that could be visualized through the IR images.

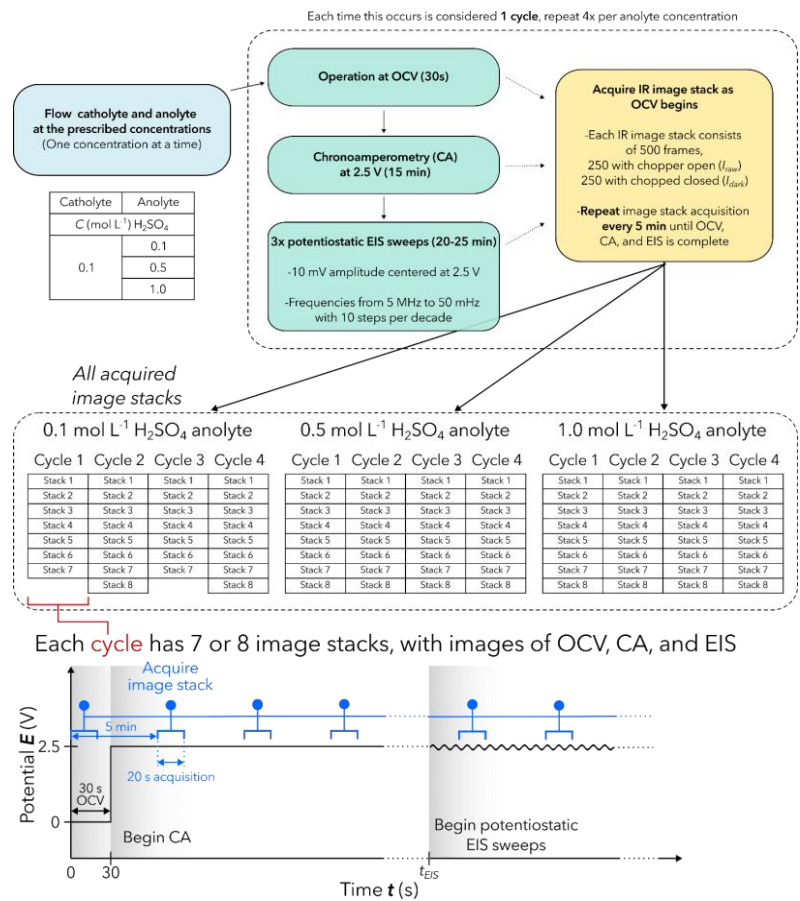

**Figure 2.** *Flowchart schematic of the experimental protocol. The number of acquired images for each cycle and anolyte concentration of the experiment is shown.*

To exaggerate this phenomenon, three anolyte concentrations (0.1 mol $L^{-1}$, 0.5 mol $L^{-1}$, and 1.0 mol $L^{-1}$) were selected and the catholyte $H_2SO_4$ concentration was held constant at 0.1 mol $L^{-1}$. A syringe pump was used to flow each reactant at 100 μL $min^{-1}$, and the temperature of the electrolyzer was held constant at 60 °C through a custom copper thermistor (Captec) that was inserted between the electrolyzer and the stage. The temperature of the copper thermistor was controlled through an externally connected proportional-integral-derivative (PID) controller. Both the reactant delivery to the electrolyzer and its temperature control were maintained for 15 min before potential was applied.

The electrolyzer operation was separated into 4 cycles per anolyte concentration for a total of 12 cycles. Each cycle consisted of open circuit voltage (OCV) operation for 30 s, then a constant potential step at 2.5 V for 15 min, followed by three potentiostatic EIS sweeps centered at 2.5 V with an amplitude of 10 mV (frequencies from 5 MHz to 50 mHz with 10 steps per decade). Immediately after beginning OCV, 500 frames of images were acquired at a frequency of 25 Hz every 5 min until the potentiostatic EIS sweeps were complete (but not presented in this work). Eight images were acquired for each cycle and concentration, with exceptions of seven images acquired for cycle 1 and cycle 3 with 0.1 mol L$^{-1}$ anolyte. A schematic describing the experimental protocol is shown in **Erreur ! Source du renvoi introuvable.**. Each 50 frame image stack acquired images alternating between $I_{raw}$ and $I_{dark}$, such that 250 frames of each were acquired per image stack.

3. Measurements of the membrane hydration

Acquired images must be processed using a rewritten Beer Lambert law due to the experimental setup. As the IR camera captures both IR light and the proper emission of the sample, the proper emission component must be removed to determine the amount of IR light attenuated by the sample. The proper emission of the sample can be determined through simple means, either by turning the beam off or blocking it to capture dark-field images, such as when the chopper is closed. Consequently, the revised Beer Lambert law can be rewritten as such:

$$l_w(x,y,t) = -\frac{1}{\mu_w}\log\left(\frac{I(x,y,t) - I_{dark}(x,y,t)}{I_0(x,y,t) - I_{0,dark}(x,y,t)}\right) \quad (1)$$

where $I$ is the operating image acquired while potential is applied to the electrolyzer [-], $I_0$ is the reference image acquired while the electrolyzer is operated at OCV [-], $I_{dark}$ is the dark-field image acquired with the chopper closed [-], $\mu_w$ is the attenuation coefficient for water at a wavelength of 4 µm [µm-1], and $l_w$ is the change in the length of water through which the beam is attenuated from the reference state to operating state [µm]. As images were acquired alternating between $I$ (previously denoted as $I_{raw}$) and $I_{dark}$, the subsequent $I_{dark}$ frame was subtracted from the previous frame of $I$ when **Erreur ! Source du renvoi introuvable.**(1) was applied. A water attenuation calibration was performed using the experimental setup detailed in [10] to calculate $\mu_w$ at the specified wavelength, which was determined to be 5.9 x·10$^{-3}$ µm$^{-1}$. The change in the length of water in the beam path can be converted to the change in membrane hydration ($\overline{\lambda_{H_2O}}$) [-] through the following relation:

$$\overline{\lambda_{H_2O}}(x,y,t) = \frac{C_{w,px^2}(x,y,t)}{C_{SO_3H,px^2}} \quad (2)$$

where $C_{w,px^2}$ is the concentration of water in the PEM per square pixel [mol px$^{-2}$] and $C_{SO_3H,px^2}$ is the equivalent per sulfonate site in the PEM [mol px$^{-2}$]. Each concentration component can be determined through the following equations:

$$C_{w,px^2}(x,y,t) = \frac{l_w(x,y,t)\rho_w res_{px}}{M_w} \quad (3)$$

$$C_{SO_3H,px^2} = TAC\,\rho_{PEM}\,t_{PEM}\,res_{px} \quad (4)$$

where $\rho_w$ and $\rho_{PEM}$ are the densities for water and the PEM [g cm$^{-3}$], respectively, $res_{px}$ is the resolution of a square pixel [µm2 px$^{-2}$], $M_w$ is the molar mass of water [g mol$^{-1}$], $TAC$ is the total acid capacity of the PEM [g molSO$_3$H$^{-1}$],[11] and $t_{PEM}$ is the thickness of the PEM [µm].

## 4. Results

### 4.1 Segmenting images into local regions

Here, the membrane hydration in local regions of the same operating electrolyzer was visualized through processed images using combination of both raw and $\overline{\lambda_{H_2O}}$ image sets. Both image sets were used for analysis to investigate changes in membrane hydration (determined from $\overline{\lambda_{H_2O}}$ images) for local regions that contained different combinations of adjacent channel wetness/dryness (determined from raw images). Water transport in the electrolyzer was then investigated for three characteristic regions (wet: anode and cathode channels both contain electrolyte, dry: anode and cathode channels both in absence of electrolyte, and hybrid: dry anode channel and wet cathode channel). In Fig 3a, each component of the electrolyzer is shown for a sample raw image with three red boxes shown that represent the areas for each characteristic region. For this analysis, only one sample image was used (1.0 mol L$^{-1}$ anolyte, cycle 1 image 7), for which the sample image was selected such that these three characteristic regions could be observed simultaneously. These same regions are labeled for the $\overline{\lambda_{H_2O}}$ image in Fig 3a, where the water gradient in these characteristic regions was used to investigate PEM water transport mechanisms. $\overline{\lambda_{H_2O,a}}$ and $\overline{\lambda_{H_2O,c}}$ from Fig 3 will be considered at the $y$ values in the red boxes from Fig 3 a and b that are closest to the respective electrodes.

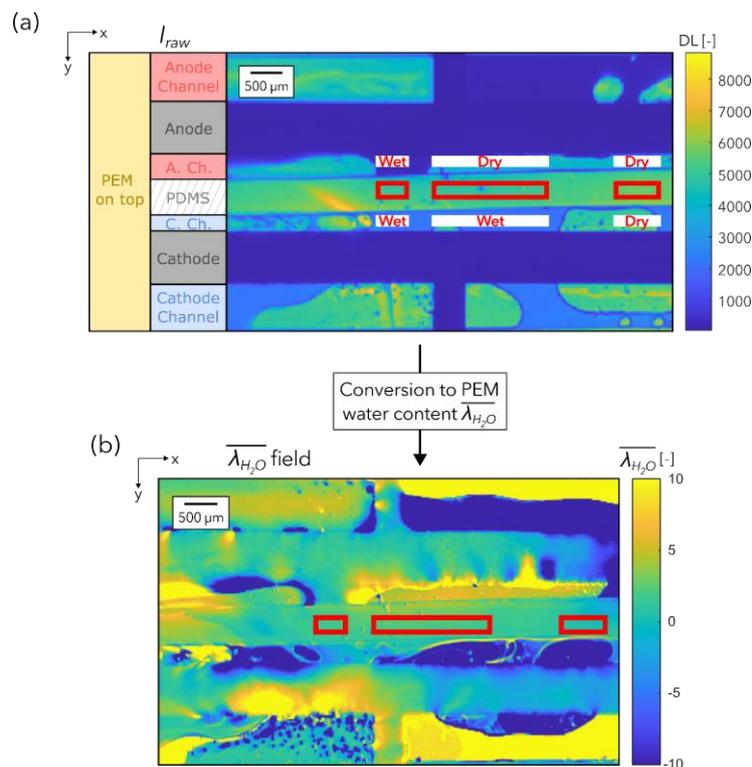

**Figure 3.** *(a) A corresponding sample image at cycle 1 with 1.0 mol L$^{-1}$ anolyte with all cell components labeled on the left, and the three local regions that are denoted by red boxes were investigated. From left to right, first is the wet region, where the adjacent channels contain the electrolyte. Second is the hybrid region, where the immediately adjacent anode channel is dry while the adjacent cathode channel is wet. Third is the dry region, where both adjacent channels are dry. (b) change in membrane hydration is characterized for the red boxes that represent each of these regions. The wavelength used is 4 µm.*

Considering the experiment with 1.0 mol L$^{-1}$ anolyte from cycle 1 image 7 and the specific regions in Fig a and b, the leftmost region has an anode channel wetness of 100 % and cathode channel wetness of 87.5 % and will be here on referred to as the wet region. Next, the middle region has an anode channel wetness of 43.0 % and cathode channel wetness of 70.6 % and will be referred to as the hybrid region, and finally the rightmost region has an anode channel wetness of 0.3 % and a cathode wetness of 38.2 % and will be referred to as the dry region.

Binarized images were averaged to visualize the movement of water in the adjacent channels where 0 corresponds to a dry channel and 1 corresponds to a wet channel. It is important to note that the influence of bubbles on the fluid dynamics in this particular electrolyzer are inefficient due to the morphology of the electrodes, and also because the cell was fabricated in absence of a porous transport layer. The smooth electrode surface combined with the lack of a porous transport layer causes low bubble contact angles that can complicate bubble removal [12,13]. The reported wetness percentages for each characteristic region may be strongly influenced by the lack of components/morphology for improving mass transport.

4.2 Effect of local channel wetness on PEM hydration

With each region defined in the previous section, changes in membrane hydration ($\overline{\lambda_{H_2O}}$) will be investigated. In each region, $\overline{\lambda_{H_2O}}$ will be compared in terms of five evenly spaced columns and rows, to observe how the adjacent channel wetness affects local PEM hydration. Each column of interest in the characteristic regions can be directly related to $\Delta\overline{\lambda_{H_2O}}$, as the column of interest can represent the path of proton transport from the anode to cathode. However, particles/imaging artifacts that interfere with imaging may also affect the $\overline{\lambda_{H_2O}}$ profile of each column. $\Delta\overline{\lambda_{H_2O}}$ values would consequently be especially sensitive to noise from these artifacts because each characteristic region only has 10 px in the y-direction. To dampen the effect of these imaging artifacts, a linear trend is assumed for $\Delta\overline{\lambda_{H_2O}}$ changes over the y-direction of the region of interest and then fitted over the 10 px length in the y-direction. For any column of interest at a given x-position, the fitted change in membrane hydration across the 10-pixels in the y-direction between electrodes can be defined:

$$\Delta\overline{\lambda_{fit}}(x) = a_1 x + a_0 \qquad (4)$$

where $a_1$ and $a_0$ are coefficients fitted for each column of interest which is shown in **Erreur ! Source du renvoi introuvable.**b as dotted fit lines. For each of the five rows of interest in the characteristic regions, the $\overline{\lambda_{H_2O}}$ profiles can be used to elaborate the effects of adjacent channel wetness in the direction of flow, and any biases in membrane hydration that may exist towards electrodes can be identified.

The wet region is shown in Fig 4, with the positions for five evenly spaced columns and rows of interest shown along the edge of the $\overline{\lambda_{H_2O}}$ field. The $\overline{\lambda_{H_2O}}$ profile for each line of interest was plotted in Fig 4b for columns and Fig 4c for rows. Considering the columns of interest, a slight increase in $\overline{\lambda_{H_2O}}$ can be seen in the direction of flow. It is observed that the values of $\Delta\overline{\lambda_{fit}}$ ranges from -0.23 to -0.089 and have an average value of -0.16, indicating a slight reduction in membrane hydration between the electrodes of the wet region from the reference state at OCV. However, each row of interest (Fig 4c) increases in $\overline{\lambda_{H_2O}}$ from values around 1.4 to 2.0 in the direction of flow, regardless of the respective row's y-position. These changes in $\overline{\lambda_{H_2O}}$ indicate that when the adjacent channels contacting the PEM are wetted and contain electrolyte, the membrane hydration will consequently increase in the direction of flow, without a bias towards either electrode.

While the membrane hydration consistently increased along the direction of flow for the wet region, in the hybrid region, changes in membrane hydration for both $\overline{\lambda_{H_2O}}$ and $\Delta\overline{\lambda_{fit}}$ followed a different trend (not presented here, but processed). In this characteristic region, column data shows that $\overline{\lambda_{H_2O}}$ is highest on the anode side (y = 1, top of Fig 4a) and lowest on the cathode side (y = 10, bottom of Fig. 4b), creating a water gradient from the anode to cathode. The corresponding $\Delta\overline{\lambda_{fit}}$ values have an average of 0.89 for the hybrid region, and in the direction of flow $\Delta\overline{\lambda_{fit}}$ decreased from 1.14 to 0.70. As the hybrid region immediately followed the wet region in the direction of flow (or by increasing x-position), the column with the smallest x-position in the hybrid region was expected to have the highest $\overline{\lambda_{H_2O}}$ values. Since the hybrid region is drier for both the anode (from 100.0 % to 43.0 % channel wetness) and cathode channels (from 87.5 % to 70.6 % channel wetness), the decreasing trend of $\overline{\lambda_{H_2O}}$ in the direction of flow was expected. Considering the rows of interest, the water gradient from the columns of interest could also be visualized, as the rows closest to the anode consistently have a higher $\overline{\lambda_{H_2O}}$ than those closest to the cathode side. Overall, the results from the hybrid region show higher PEM water uptake at the anode side from OCV, despite the anode channel wetness of 43.0 % being lower than the cathode channel wetness of 70.6 %. While the anode channel wetness is lower than that of the cathode, the change in PEM hydration was still higher on the anode side. This result supports the previous hypothesis of the higher mass transport losses being tied to the cathode for this anolyte concentration.

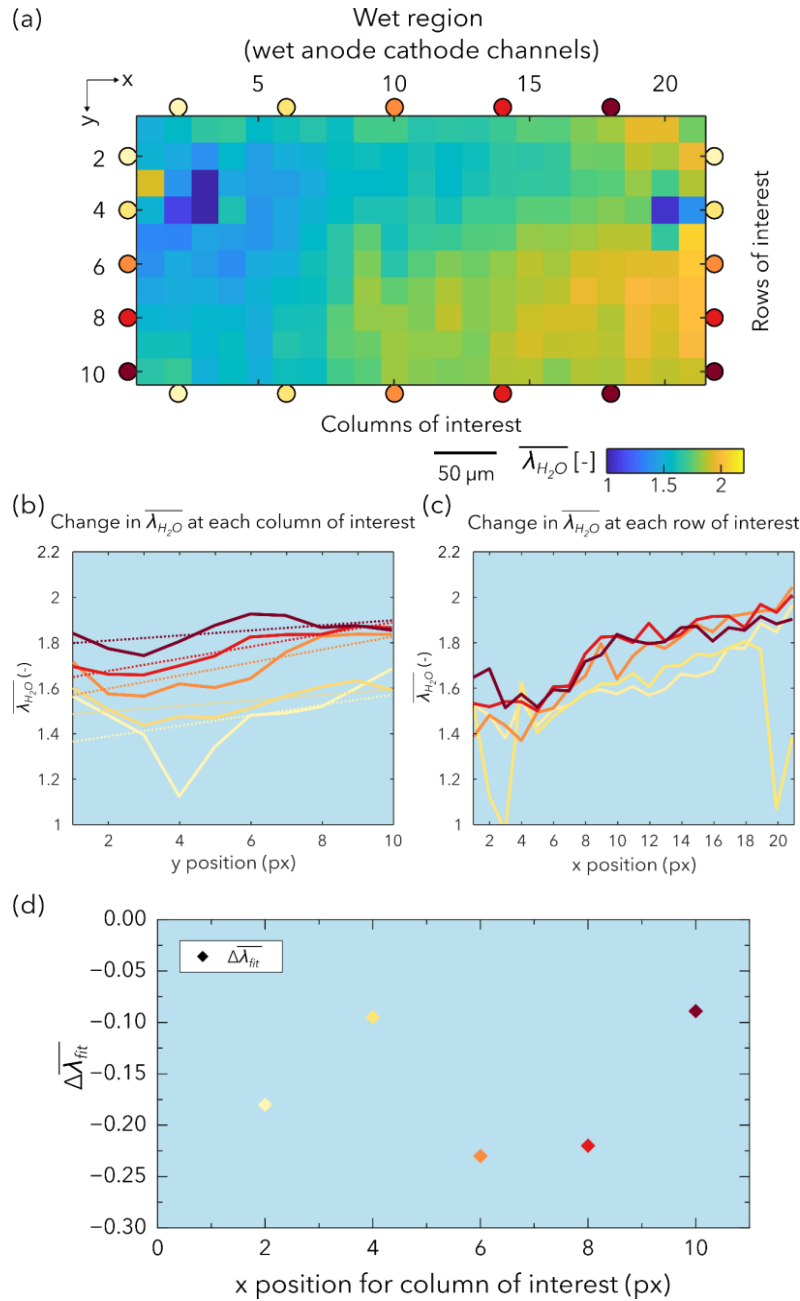

**Figure 4.** *(a) The wet region of interest with the change in membrane hydration for five (b) columns and (c) rows shown for the chip. Column and row positions are indicated by color on (a), which correspond to the plotted colors on (b) and (c). In (b), the profile of each column is fitted using (d) The values of membrane hydration with respect to each column of interest are presented. The electrolytes flow from left to right.*

4.2 Conclusion

Local imaging results were studied to investigate the effect of local channel wetness on the PEM hydration. Different characteristic regions (wet, hybrid, and dry) were observed, each selected based on the wetness of the adjacent channels. Clear gradients between anode and cathode channels were visualized, enabling the first quantification of $\lambda_{H_2O}$ fields via operando IR imaging. Water distribution through the PEM and its changes on the local and global scale showed the need for components such as PTLs to distribute water evenly within the channels. Once a microfluidic PEM electrolyzer is developed with improved water distribution, the variables affecting $\lambda_{H_2O}$ can be isolated and investigated more easily. Additionally, mass transport losses would be lower at the same conditions. In addition, the results from this work can immediately be improved in the next design iteration with the inclusion of PTLs for improved water management in the channels. The microfluidic PEM electrolyzer and IR imaging setup were developed as a platform for characterization, for which we have effectively shown its potential through the measurement of PEM hydration during electrochemical operation.